\documentclass[runningheads]{llncs}

\usepackage{graphicx,color,xcolor}
\usepackage{listings}
\usepackage{amsmath}

\usepackage{hyperref}
\usepackage{cleveref}
\usepackage{amssymb}
\usepackage{listings}
\usepackage{comment}
\usepackage{mathtools}
\usepackage[hyperref=true,backend=biber,giveninits=true,style=numeric,sorting=nyt,maxbibnames=4, maxnames=4]{biblatex}
\newcommand{\globid}{\textit{Globally}}
\newcommand{\locid}{\textit{Locally not Globally}}
\newcommand{\unid}{\textit{Non-Identifiable}}
\newcommand{\se}{\textit{Single-Experiment}}
\newcommand{\me}{\textit{Multi-Experiment}}
\begin{document}

\title{Web-based Structural Identifiability Analyzer\thanks{This work was partially supported by the NSF grants CCF-1564132, CCF-1563942, DMS-1760448,  DMS-1853650, and DMS-1853482, and by the Paris Ile-de-France Region.}}
%
%
\author{Ilia Ilmer\inst{1} \and
	Alexey Ovchinnikov\inst{2} \and
	Gleb Pogudin\inst{3}}
\authorrunning{I. Ilmer et al.}

\institute{Ph.D. Program in Computer Science, CUNY Graduate Center, New York, USA, \email{iilmer@gradcenter.cuny.edu}\\ \and Department of Mathematics, CUNY Queens College and Ph.D. Programs in Mathematics and Computer Science, CUNY Graduate Center, New York, USA;\@
	\email{aovchinnikov@qc.cuny.edu}\\ \and
	LIX, CNRS, École Polytechnique, Institute Polytechnique de Paris, France\\
	\email{gleb.pogudin@polytechnique.edu}}

\maketitle              

\begin{abstract}
	Parameter identifiability describes whether, for a given differential model, one can determine parameter values from model equations. Knowing global or local identifiability properties allows construction of better practical experiments to identify parameters from experimental data.
	In this work, we present a web-based software tool that allows to answer specific identifiability queries. Concretely, our toolbox can determine identifiability of individual parameters of the model and also provide all functions of parameters that are identifiable (also called identifiable combinations) from single or multiple experiments. The program is freely available at \url{https://maple.cloud/app/6509768948056064}.
\end{abstract}

\section{Introduction and Related Work}

A parameter is said to be structurally \emph{globally} identifiable if, given the input and output of the experiment, one can uniquely recover the parameter's value in the generic case.
If the recovered value is not unique but comes from a finite collection, then we say that such a parameter is \emph{locally} identifiable. Otherwise, the parameter is called \emph{non-identifiable}.
In the latter case, one wonders if there is a function of that parameter that is identifiable. This is useful in several ways, for instance, it can mitigate the issue of non-identifiability of some parameters~\cite{ovchinnikov2020computing}.

There is a variety of installable packages that deal with parameter identifiability, see, for instance~\cite{bellu2007daisy,chics2011genssi,ligon2018genssi,saccomani2019daisy2,villaverde2016structural}.
For a more detailed overview of these, see~\cite{comparison,hong2020global} and references therein. A general overview of solving parameter identifiability problems was presented, for instance, in \cite{miao2011identifiability,raue2014comparison,villaverde2019observability}.
Among the available identifiability software,
SIAN~\cite{SIAN} written in {\sc Maple}\footnote{For a Julia implementation, see \url{https://github.com/alexeyovchinnikov/SIAN-Julia}} is typically the fastest one for assessing global identifiability of individual parameters (see, e.g., \cite[Table~1]{SIAN}).
In the case of the lack of identifiability, one may want to find which functions of the parameters are identifiable.
For this task, DAISY software~\cite{bellu2007daisy} implemented in Reduce can be used (under some assumptions, see~\cite[Remark~3]{Saccomani2003} and~\cite{ovchinnikov2020computing}).
This state of affairs may be inconvenient for the user because
\begin{enumerate}
	\item[1)] the features of interest are scattered among different packages;
	\item[2)] packages may require proprietary ({\sc Maple}) or less popular (Reduce) software and may not be available for commonly used OS (DAISY is not available for the UNIX-type systems);
	\item[3)] finally, the packages should be installed.
\end{enumerate}

These issues have been partially addressed by a web-based tool called {\sc COMBOS}~\cite{meshkat2014finding} (and its recent refinement {\sc COMBOS} 2 for linear systems~\cite{kalami2020combos2}).
However, the backend algorithm appears to be less efficient than SIAN~\cite[Table~1]{SIAN}, and it relies on the same assumption on the input model as DAISY.\@

Our main contribution is a web-based toolbox hosted on Maple Cloud for assessing structural identifiability built upon SIAN and recent software for computing identifiable functions of parameters~\cite{ovchinnikov2020computing} which uses the Boulier's {\sc BLAD} software package~\cite{boulier2014blad} incorporated into the {\sc Maple}'s {\tt Differential Algebra} package.
The key features are
\begin{enumerate}
	\item[1)] \emph{efficiency.} We use SIAN for assessing identifiability of individual parameters efficiently.
		For computing identifiable functions, we use the code from~\cite{ovchinnikov2020computing} which we speed up by exploiting the results of the computation performed by SIAN.\@
	\item[2)] \emph{versatility.} The toolbox allows assessing local and global identifiability of the parameters and initial conditions and compute the identifiable functions in parameter both in the single- and multi-experiment setup.
		We do not make any assumptions on the input system unlike DAISY or COMBOS.\@
	\item[3)] \emph{availability.} The toolbox is a web app, so it can be used in a browser in one click and does not require installing anything.
\end{enumerate}

In \Cref{sec:use_cases}, we outline several scenarios in which our application is essential for assessing identifiability of parameters and parameter combinations. We also illustrate the speedup achievable using output from each of its parts. The web-application\footnote{The {\sc Maple} implementations of each underlying algorithm are available on GitHub at \url{https://github.com/pogudingleb/SIAN} and \url{https://github.com/pogudingleb/AllIdentifiableFunctions}.} can be used at \url{https://maple.cloud/app/6509768948056064} and is also available for download.

\section{Input-output specification}
Let us define the specific form of state-space input ODE that our application accepts.
\begin{definition}[Model in the state-space form]
	\label{def1}
	\normalfont{} A model in \emph{the state-space form} accepted by the application is a system
	\[ \mathbf{\Sigma} := \begin{cases}
			\mathbf{x}'   & =\mathbf{f}(\mathbf{x}, \boldsymbol\mu, \mathbf{u}), \\
			\mathbf{y}    & =\mathbf{g}(\mathbf{x}, \boldsymbol\mu, \mathbf{u}), \\
			\mathbf{x}(0) & =\mathbf{x}^\ast,
		\end{cases}\]
	where \(\mathbf{f}=(f_1,\dots, f_n)\) and \(\mathbf{g}=(g_1,\dots,g_n)\) with \(f_i=f_i(\mathbf{x}, \boldsymbol\mu, \mathbf{u})\), \(g_i=g_i(\mathbf{x}, \boldsymbol\mu, \mathbf{u})\) are rational functions over the field of complex numbers \(\mathbb{C}\).

	The vector \(\mathbf{x}=(x_1,\dots,x_n)\) represents the time-dependent state variables and \(\mathbf{x}'\) represents the derivative. The vector-function \(\mathbf{u}=(u_1,\dots,u_s)\) represents the input variable. The \(m\)-vector \(\mathbf{y}=(y_1,\dots,y_n)\) represents the output variables. The vector \(\boldsymbol\mu=(\mu_1,\dots,\mu_{\lambda})\) represents the parameters and \(\mathbf{x}^\ast=(x_1^\ast,\dots,x_n^\ast)\) defines initial conditions of the model.
\end{definition}

Below we specify the input format and possible outputs of our toolbox. Note that while used in descriptions below, some outputs, such as number of solutions for each parameter, are not listed here for brevity. The app also provides additional logs for debugging purposes. In \Cref{appendix}, we provide more specification examples.

\begin{align*}
	 & \begin{array}{rl}
		\textbf{In:} & \text{A model in state-space form, see \Cref{def1}.}
	\end{array} \\
	 & \begin{array}{rll}                                                                                             \\
		\textbf{Out:} & \globid: & \text{Globally identifiable parameters, that is}                       \\
		              &          & \text{ones uniquely recoverable for a given system.}                   \\
		              & \locid:  & \text{Locally but not globally identifiable}                           \\
		              &          & \text{parameters, with finitely many recoverable values.}              \\
		              & \unid:   & \text{Non-identifiable parameters, these can have}                     \\
		              &          & \text{infinitely many values.}                                         \\
		              & \se:     & \text{Single-Experiment identifiable functions of}                     \\
		              &          & \text{parameters, i.e. identifiable from \(k\leq 1\) experiments.}     \\
		              & \me:     & \text{Multi-Experiment identifiable functions of}                      \\
		              &          & \text{parameters, i.e.\ identifiable from \(k\leq\beta\) experiments.} \\
		              & \beta:   & \text{Bound on the number of experiments.}
	\end{array}
\end{align*}
Note that the single- and multi-experiment identifiable combinations returned by the app generate \emph{all} single- and multi-experiment functions of parameters, respectively.
We return them in the algebraically simplified form. In addition, the app reports number of solutions per each globally or locally identifiable parameter, which is not explicitly reflected here due to space limitations.


\section{Use Cases for Structural Identifiability Toolbox}
\label{sec:use_cases}
\subsection{Globally Identifiable Example (two-species competition model)}

Let us consider a simple two-species competition model based with logistic growth in homogeneous environment and assume that we are interested in identifiability properties of all parameters and initial conditions:
\[
	\begin{cases}
		x_1'  = r_1x_1\left(1 - \tfrac{x_1+x_2}{k_1}\right), \\
		x_2' = r_2x_2\left(1 - \tfrac{x_1+x_2}{k_2}\right),  \\
		y_1   = x_1,~~y_2 = x_2
	\end{cases}
\]
with population densities \(x_1,x_2\) being time-dependent state variables, and intrinsic growth rates \(r_1,r_2\) and carrying capacities \(k_1,k_2\) being constant. To run the toolbox for this system, we would write the following into the input field:
\begin{align*}
	 & \begin{array}{r@{}l}
		\textbf{In:~~} & \texttt{diff(x1(t),t) = r1*x1(t)*(1 - (x1(t) + x2(t))/k1);} \\
		               & \texttt{diff(x2(t),t) = r2*x2(t)*(1 - (x1(t) + x2(t))/k2);} \\
		               & \texttt{y1(t) = x1(t);}                                     \\
		               & \texttt{y2(t) = x2(t)}
	\end{array} \\
	 & \begin{array}{r@{}ll}
		\textbf{Out:~~} & \globid: & \texttt{[x1(0), x2(0), r1, r2, k1, k2]} \\
		                & \locid:  & \texttt{[]}                             \\
		                & \unid:   & \texttt{[]}
	\end{array}
\end{align*}
To determine the identifiability for this model, we keep default ``Check global/local identifiability'' and ``Print Number of Solutions'' options on. After entering the system and running the application, the output field contains the results. In this model, all parameters and initial conditions are globally (and locally) identifiable. One can now proceed to data collection and further experiments.

\subsection{Locally Identifiable Model (SIRS model with forcing)}
\label{sirsforced}
Consider an example of a seasonal epidemic model with a periodic forcing term:
\[
	\begin{cases}
		s'=\mu - \mu s - b_0(1 + b_1x_1)i\cdot s+ g\cdot r, \\
		i'= b_0(1 + b_1x_1)i\cdot s - (\nu+\mu) i,          \\
		r' = \nu i - (\mu + g)r,                            \\
		x_1' = -Mx_2,                                       \\
		x_2' = Mx_1,                                        \\
		y_1 = i,\ \ y_2 = r.
	\end{cases}
\]
The model is taken from~\cite{capistran2009parameter} and is built into the application as one of the illustrating examples.
Assume that we are interested in identifiability of parameters of this model.
Without changing default settings, running the application yields the result of \(b_1,x_1(0),x_2(0)\) being unidentifiable, and \(b_0,g,\mu,\nu,s(0),i(0),r(0)\) as globally identifiable.
At the same time, we observe that \(M\) which defines oscillation of the term \(x_1\) is the only parameter identifiable locally, not globally.
By checking the number of solutions, we see that only two can be found for \(M\) with probability \(p=0.99\). Since \(M\) represents the oscillation frequency, it is assumed to be positive in practice, hence globally identifiable.
Note that we only needed a single section of the app and the result has been obtained in about 7.2 seconds.

\subsection{Identifiable Combination of Non-Identifiable Parameters (tumor targeting)}
\label{tumor}
In this example, we consider system 3 from~\cite[Section~3]{saccomani2010examples} with unknown initial conditions. The example describes a  compartmental model describing tumor targeting with antibodies, see~\cite{thomas1989effect}. To arrive at the system below, we suppose equations (B) and (D) are identically zero and that \(\frac{5V36}{V3}=1\). The functions \(x_i,i=1,\dots,5\) represent concentrations, \(k_i,i=3,\dots,7\) and \(a, b, d\) represent rate constants.
\[
	\begin{cases}
		x_1' = -(k_3 + k_7)x_1 + k_4x_2,                                              \\
		x_2' = k_3x_1 - (k_4 + (a + bd)k_5)x_2 + k_6(x_3 + x_4)  + k_5x_2(x_3 + x_4), \\
		x_3' = ak_5x_2 - k_6x_3 - k_5x_2x_3,                                          \\
		x_4' = bdk_5x_2 - k_6x_4 - k_5x_2x_4,                                         \\
		x_5' = k_7x_1,                                                                \\
		y_1 = x_5.
	\end{cases}
\]

For this model, after computing identifiability properties using SIAN, we observe that everything except parameters \(a,b,d,x_3(0),x_4(0)\) is globally or locally identifiable. To investigate further, we consider computation with ``Compute Identifiable Combinations'' option turned on. Running the program with this additional setting, we see that while parameters \(a,b,d\) are not identifiable, their combination \(a+bd\) can be identified from at most one experiment. This is especially beneficial since one can connect the meaning of expression \(a+bd\) to the overall biological sense of the model's underlying phenomenon. For instance, in the original paper~\cite{thomas1989effect}, constant \(a\) and a product \(bd\) may be attributed to total binding sites on normal tissue and number of binding sites on tumor making \(a+bd\) the total number of binding sites in the system.
Further, one could apply a substitution of the form \(\widehat{x}_3 = x_3+x_4, \widehat{p}=a+bd\) so that in the new system we only have equations for \(x_1, x_2, \widehat{x}_3, x_5\) and the parameter combination \(a+bd\) will now be globally identifiable as a parameter \( \widehat{p}\).

\subsection{System with a Non-Identifiable Parameter (Lotka-Volterra model)}
\label{LV}
Let us consider the following Lotka-Volterra model
\begin{equation}\label{eq:LV}
	\begin{cases}
		x_1'  = ax_1 - bx_1x_2  \\
		x_2'  = -cx_2 + dx_1x_2 \\
		y     = x_1
	\end{cases}
\end{equation}

By running the application for~\eqref{eq:LV} using only SIAN, we see that parameter \(b\) and initial condition \(x_2(0)\) are non-identifiable, and the parameters \(a, b, d\) and the initial condition \(x_1(0)\) are globally identifiable.
Furthermore, since \(a\) is identifiable and \(x_1\) is observed, from the first equation we conclude that \(bx_2(0) = a - x_1'(0)/x_1(0)\) is identifiable. This implies that we have an output-preserving scaling transformation \(b \to \lambda b\), \(x_2 \to x_2/\lambda\). Therefore, the reparametrization \(\hat{x}_2 := bx_2\) makes the model globally identifiable.

\subsection{Refining Multi-Experiment Identifiability Bound\\(slow-fast ambiguity in a chemical reaction network)}
\label{slowfast}
Consider the following system:
\[
	\begin{cases}
		x_A' = -k_1x_A,                  \\
		x_B' = k_1x_A - k_2x_B,          \\
		x_C' = k_2x_B,                   \\
		e_A' = e_C'=0,                   \\
		y_1  = e_Ax_A + e_Bx_B + e_Cx_C, \\
		y_2  = x_C, ~y_3  = e_A, ~y_4  = e_C.
	\end{cases}
\]

This model is based on a kinetic reaction \(A\xrightarrow[]{k_1} B\xrightarrow[]{k_2} C\) from~\cite{vajda1988identifiability} and has an extra output equation \(y_2\). The functions \(x_A,~x_B,~x_C\) are concentrations and \(e_A,~e_B\) with constant \(e_C\) represent molar extinction coefficients. In addition, parameters include unknown rate coefficients \(k_1,~k_2\). The application reports global identifiability for \(x_C(0),~e_A(0),~e_C(0)\) and local identifiability for everything else.

It is then of interest to check identifiable parameter combinations. The app reports single-experiment identifiability for \(k_1k_2,~k_1+k_2\). This implies that the parameters \(k_1\) and \(k_2\) are identifiable up to a permutation, so it is possible to infer the reaction rates from an experiment but not which rate corresponds to which reaction.
Interestingly, the app reports that \(e_B,~k_1,~k_2\) become globally identifiable if one performs at most 3 experiments.
Can we do better? To answer this, we turn on the option ``Try to Refine Bound'' with default number of refining attempts being 4. As a result, the app reports a new bound for the number of experiments being 2.

Let us illustrate this point in another way. Recall that we can tell SIAN to consider multiple copies of the system when analyzing identifiability. In this mode, SIAN does not output initial conditions for brevity. We observed that the refined bound for parameters \(e_B,k_1,k_2\) was 2. If we set the ``Number of experiments (copies of the input system)'' to 2, SIAN yields global identifiability of \(e_B,k_1,k_2\), which verifies our earlier finding. Moreover, turning off ``Attempt Bypass using SIAN'' option in the search for combinations, we observe that the application still returns \(e_B,k_1,k_2\) as identifiable with 3 experiments, however, single experiment check overwrites this result, yielding bound of 1.

\section*{Acknowledgements}
We are grateful to Joseph DiStefano III, J\"urgen Gerhard, John May, Maria Pia Saccomani, and Eduardo Sontag for fruitful discussions, useful feedback, and technical assistance.

%
%
%
\printbibliography{}
\pagebreak
\appendix
\section{Details on the underlying algorithms}

The application solves two problems: identifiability properties of individual parameters and that of combinations (functions) of parameters. Note that we return generators of the field of \emph{all} identifiable functions.

The input for both problems follows the same structure where we pass a collection of ODEs and output functions. For querying identifiability of individual parameters and initial conditions we use SIAN~\cite{SIAN}. In short, it expresses the Taylor coefficients of output functions in terms of the initial conditions and parameters and checks whether the parameters or initial conditions of interest can be expressed via these coefficients. For better efficiency, this is checked for a randomly sampled solution of the system. The probability that such a solution will exhibit the generic behavior is quantified in~\cite{hong2020global}.
Therefore, the overall algorithm is randomized Monte Carlo, that is the result is guaranteed correct with user-specified probability \(p\).

To answer the question on identifiability of parameter functions we take advantage of work~\cite{ovchinnikov2020computing}. Note that our application distinguishes single- and multi-experiment identifiable combinations as opposed to existing methods for identifiable combination queries. The latter is equivalent to having multiple copies of original ODE system sharing the parameters, outputs, and inputs. We also provide a bound on the number of experiments which can be refined by changing ordering of variables in the underlying algorithm.

We compute the input-output equations, that is, differential equations relating inputs, outputs, and parameters of the differential model. Identifiable functions of parameters are then extracted from the coefficients of these equations using methods of differential algebra  and computational algebraic geometry, including Gr\"obner basis computation. To minimize the computational overhead, we take advantage of the Gr\"obner walk procedure, by changing the order from total degree reverse to pure lexicographic. This algorithm is deterministic or a Monte Carlo probabilistic, depending on how/which of the Gr\"obner basis implementation is used.

To achieve maximal speed of computation without compromising the functionality of the application, we take advantage of the fact that SIAN is typically faster than the algorithm from~\cite{ovchinnikov2020computing} and its output can be sometimes used to obtain the output of~\cite{ovchinnikov2020computing} without further computation. More precisely, if all parameters are reported as globally identifiable with probability \(p\), then, with the same probability, we report these parameters as their own identifiable combinations and an example of this is presented in \Cref{crn}.

With the current implementation, the application does not support specifying initial conditions, however this functionality is planned for future versions.
\begin{figure}
	\centering
	\includegraphics[width=0.9\linewidth]{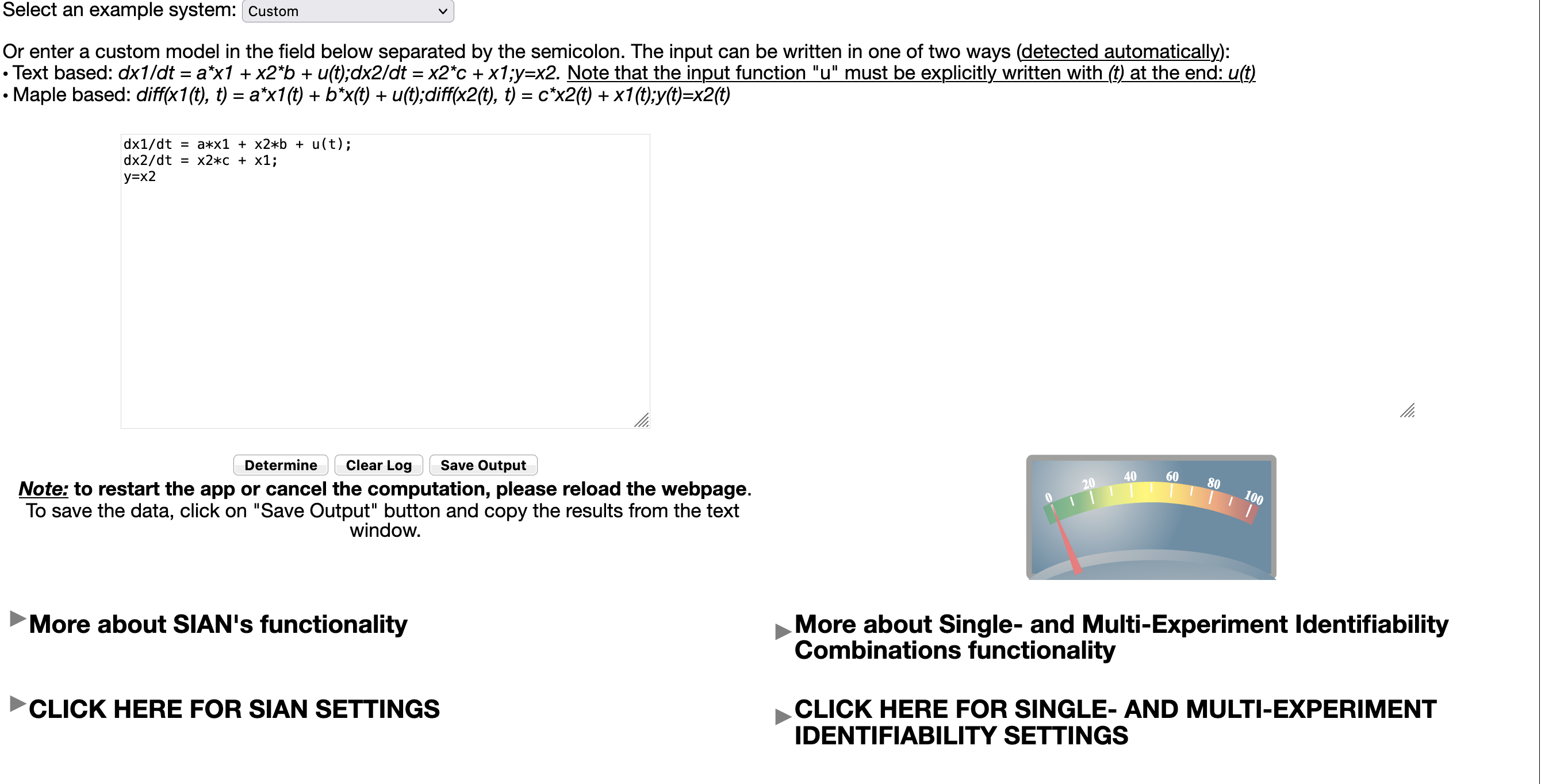}
	\caption{Main view of the application. The dial on the right side indicates whether an app is running. The arrows are clickable and show additional settings for each section of the program as well as documentation.}
\end{figure}
\begin{figure}
	\centering
	\includegraphics[width=0.9\linewidth]{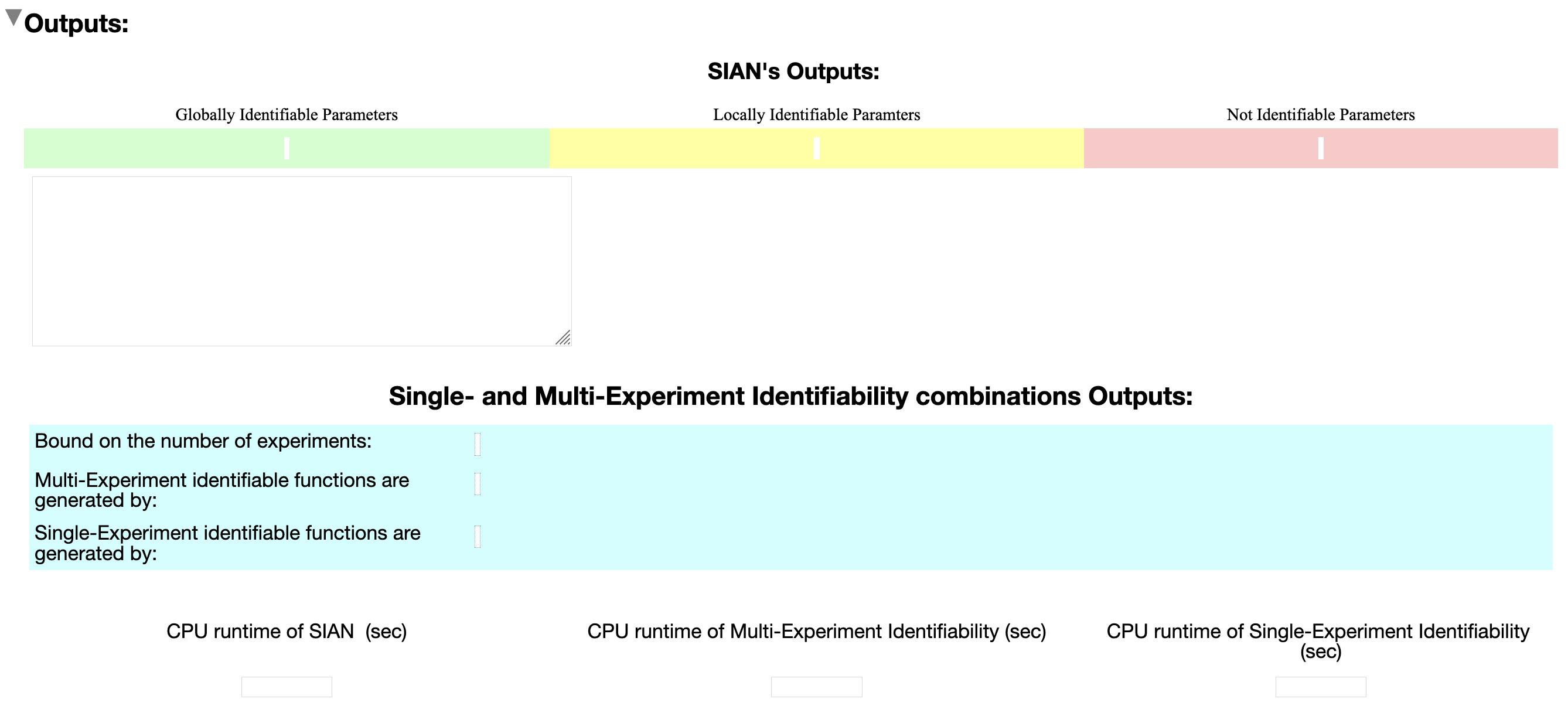}
	\caption{Output fields of the application. We present individual parameters' results and identifiable combinations with the bound separately. The white field on the left displays number of solutions per identifiable parameter.}
\end{figure}

\section{Systems in Structural Identifiability Toolbox Input Form}
\label{appendix}
Below we present input and output form for examples discussed in this paper. The input is shown in the {\sc Maple} syntax form. The toolbox also supports a different input format:
\begin{align*}
	 & \texttt{dx1/dt = a*x1 + x2*b + u(t);} \\
	 & \texttt{dx2/dt = x2*c + x1;}          \\
	 & \texttt{y = x2}
\end{align*}
where inputs {\tt u(t)} are required to have argument written explicitly.
\subsection{Example from \Cref{sirsforced}}
\begin{align*}
	 & \begin{array}{r@{}l}
		\textbf{In:~~} & \texttt{diff(s(t), t) = mu - mu*s(t) - b0*(1 + b1*x1(t))*i(t)*s(t)}  \\
		               & \texttt{+ g(t)*r(t);}                                                \\
		               & \texttt{diff(i(t), t) = b0*(1 + b1*x1(t))*i(t)*s(t) - (nu+mu)*i(t);} \\
		               & \texttt{diff(r(t), t) = nu*i(t) - (mu + g)*r(t);}                    \\
		               & \texttt{diff(x1(t), d) = -M*x2(t);}                                  \\
		               & \texttt{diff(x2(t), d) = M*x1(t);}                                   \\
		               & \texttt{y1(t) = i(t);}                                               \\
		               & \texttt{y2(t) = r(t)}                                                \\
	\end{array} \\
	 & \begin{array}{r@{}ll}
		\textbf{Out:~~} & \globid: & \texttt{[b0, g, mu, nu, s(0), i(0), r(0)]} \\
		                & \locid:  & \texttt{[M]}                               \\
		                & \unid:   & \texttt{[b1, x1(0), x2(0)]}
	\end{array}
\end{align*}
\subsection{Example from \Cref{tumor}}
\begin{align*}
	 & \begin{array}{r@{}l}
		\textbf{In:~~} & \texttt{diff(x1(t),t) = -(k3 + k7)*x1(t) + k4*x2(t);}               \\
		               & \texttt{diff(x2(t), t) = k3*x1(t) - (k4 + (a + b*d)*k5)*x2(t)}      \\
		               & \texttt{+ k6*(x3(t) + x4(t)) + k5*x2(t)*(x3(t) + x4(t));}           \\
		               & \texttt{diff(x3(t), t) = a*k5*x2(t) - k6*x3(t) - k5*x2(t)*x3(t);}   \\
		               & \texttt{diff(x4(t), t) = b*d*k5*x2(t) - k6*x4(t) - k5*x2(t)*x4(t);} \\
		               & \texttt{diff(x5(t), t) = k7*x1(t);}                                 \\
		               & \texttt{y1(t) = x5(t)}                                              \\
	\end{array} \\
	 & \begin{array}{r@{}ll}
		\textbf{Out:~~} & \globid: & \texttt{[k3, k4, k5, k6, k7, x1(0), x2(0), x5(0)]} \\
		                & \locid:  & \texttt{[]}                                        \\
		                & \unid:   & \texttt{[a, b, d, x3(0), x4(0)]}                   \\
		                & \se:     & \texttt{[k3, k4, k6, k7, k5/k7, bd+a]}             \\
		                & \me:     & \texttt{[k3, k4, k6, k7, k5/k7, bd+a]}             \\
		                & \beta=   & 1
	\end{array}
\end{align*}
\subsection{Example from \Cref{LV}}
\begin{align*}
	 & \begin{array}{r@{}l}
		\textbf{In:~~} & \texttt{diff(x1(t), t) = a*x1(t) - b*x1(t)*x2(t);}  \\
		               & \texttt{diff(x2(t), t) = -c*x2(t) + d*x1(t)*x2(t);} \\
		               & \texttt{y(t) = x1(t)}                               \\
	\end{array} \\
	 & \begin{array}{r@{}ll}
		\textbf{Out:~~} & \globid: & \texttt{[a, c, d, x1(0)]} \\
		                & \locid:  & \texttt{[]}               \\
		                & \unid:   & \texttt{[b, x2(0)]}       \\
		                & \se:     & \texttt{[a, c, d]}        \\
		                & \me:     & \texttt{[a, c, d]}        \\
		                & \beta=   & 1
	\end{array}
\end{align*}
\subsection{Example from \Cref{slowfast}}
\begin{align*}
	 & \begin{array}{r@{}l}
		\textbf{In:~~} & \texttt{diff(xA(t), t) = -k1*xA(t);}                   \\
		               & \texttt{diff(xB(t), t) = k1*xA(t) - k2*xB(t);}         \\
		               & \texttt{diff(xC(t), t) = k2*xB(t);}                    \\
		               & \texttt{diff(eA(t), t) = 0;}                           \\
		               & \texttt{diff(eC(t), t) = 0;}                           \\
		               & \texttt{y1(t) = eA(t)*xA(t) + eB*xB(t) + eC(t)*xC(t);} \\
		               & \texttt{y2(t) = xC(t);}                                \\
		               & \texttt{y3(t) = eA(t);}                                \\
		               & \texttt{y4(t) = eC(t)}                                 \\
	\end{array} \\
	 & \begin{array}{r@{}ll}
		\textbf{Out:~~} & \globid: & \texttt{[xC(0), eA(0), eC(0)]}      \\
		                & \locid:  & \texttt{[eB, k1, k2, xA(0), xB(0)]} \\
		                & \unid:   & \texttt{[]}                         \\
		                & \se=     & \texttt{[k1k2, k1+k2]}              \\
		                & \me=     & \texttt{[eB, k1, k2]}               \\
		                & \beta=   & 3
	\end{array}
\end{align*}
\subsection{Example of speedup with Bypasses}
\label{crn}
Consider the system from~\cite{ChemicalReactionNetwork}
\[
	\begin{cases}
		x_1' =-k_1x_1x_2 + k_2x_4 + k_4x_6, \\
		x_2' = k_1x_1x_2 + k_2x_4 + k_3x_4, \\
		x_3' = k_3x_4 + k_5x_6 - k_6x_3x_5, \\
		x_4' = k_1x_1x_2 - k_2x_4 - k_3x_4, \\
		x_5' = k_4x_6 + k_5x_6 - k_6x_3x_5, \\
		x_6' =-k_4x_6 - k_5x_6 + k_6x_3x_5, \\
		y_1 = x_3,                          \\
		y2 = x_2.
	\end{cases}
\]

This is an example of a mixed-mechanism network, where the state functions \(x_i(t),i=1,\dots,6\) are concentrations and the parameters \(k_i,i=1,\dots,6\) are rate constants.
The app returns global and local identifiability for all parameter in under 4 seconds. This is used to conclude that multi-experiment identifiable combinations with the bound of 1 are parameters themselves. If we turn off the ``Attempt Bypass'' function, the multi-experiment identifiable combinations \(k_1,k_3,k_5,k_6,\frac{-k_2k_4 + k_3k_5}{k_2+k_3},k_2-k_3\) with bound 1 are returned in 433 seconds.
The input form for this example is presented below
\begin{align*}
	 & \begin{array}{r@{}l}
		\textbf{In:~} & \texttt{diff(x1(t), t) =-k1*x1(t)*x2(t) + k2*x4(t) + k4*x6(t);} \\
		              & \texttt{diff(x2(t), t) = k1*x1(t)*x2(t) + k2*x4(t) + k3*x4(t);} \\
		              & \texttt{diff(x3(t), t) = k3*x4(t) + k5*x6(t) - k6*x3(t)*x5(t);} \\
		              & \texttt{diff(x4(t), t) = k1*x1(t)*x2(t) - k2*x4(t) - k3*x4(t);} \\
		              & \texttt{diff(x5(t), t) = k4*x6(t) + k5*x6(t) - k6*x3(t)*x5(t);} \\
		              & \texttt{diff(x6(t), t) =-k4*x6(t) - k5*x6 (t)+ k6*x3(t)*x5(t);} \\
		              & \texttt{y1(t) = x3(t);}                                         \\
		              & \texttt{y2(t) = x2(t)}                                          \\
	\end{array} \\
	 & \begin{array}{r@{}ll}
		\textbf{Out:~~} & \globid: & \texttt{[k1, k2, k3, k4, k5, k6,} \\
		                &          & \texttt{x1(0), x2(0), x3(0),}     \\
		                &          & \texttt{x4(0), x5(0), x6(0)]}     \\
		                & \locid:  & \texttt{[]}                       \\
		                & \unid:   & \texttt{[]}                       \\
		                & \se:     & \texttt{[k1, k2, k3, k4, k5, k6]} \\
		                & \me:     & \texttt{[k1, k2, k3, k4, k5, k6]} \\
		                & \beta=   & 1
	\end{array}
\end{align*}
\end{document}